\documentclass[12pt]{article}
\usepackage{amsfonts}
\usepackage{cite}

\newcommand{\no}{\nonumber}
\newcommand{\Eqn}[1]{&\hspace{-0.5em}#1\hspace{-0.5em}&}

\newcommand{\tfrac}[2]{{\textstyle\frac{#1}{#2}}}

\newcommand{\tB}{{\tilde{B}}}
\newcommand{\tj}{{\tilde{\jmath}}}
\newcommand{\tk}{{\tilde{k}}}
\newcommand{\tu}{{\tilde{u}}}
\newcommand{\tv}{{\tilde{v}}}
\newcommand{\tx}{{\tilde{x}}}
\newcommand{\trho}{{\tilde{\rho}}}
\newcommand{\tPhi}{{\tilde{\Phi}}}

\newcommand{\hK}{{\hat{K}}}
\newcommand{\hH}{{\hat{H}}}
\newcommand{\hR}{{\hat{R}}}
\newcommand{\hS}{{\hat{S}}}

\newcommand{\cK}{N}

\newcommand{\cL}{{L_{\rm ex}}}
\newcommand{\cu}{w}
\newcommand{\cx}{y}

\newcommand{\Bex}{{B_{\rm ex}}}

\newcommand{\bbR}{{\mathbb R}}

\newcommand{\alg}[1]{\mathfrak{#1}}
\newcommand{\vpint}{\!\makebox[0pt][l]{\hspace{3.6pt}$-$}\int}

\renewcommand{\thesection}
  {\arabic{section}.\hspace{-.5em}}

\makeatletter
\renewcommand\section{\@startsection{section}{2}{\z@}%
                         {-3.25ex\@plus -1ex \@minus -.2ex}%
                         {1.5ex \@plus .2ex}%
                         {\normalfont\large\bfseries\mathversion{bold}}}
\renewcommand\subsection{\@startsection{subsection}{3}{\z@}%
                         {-3.25ex\@plus -1ex \@minus -.2ex}%
                         {1.5ex \@plus .2ex}%
                         {\normalfont\normalsize\bfseries\mathversion{bold}}}
\makeatother

\makeatletter \@addtoreset{equation}{section} \makeatother
\renewcommand{\theequation}{\arabic{section}.\arabic{equation}}

\makeatletter
\renewcommand{\appendix}{
  \renewcommand{\thesection}{Appendix \Alph{section}.\hspace{-.5em}}
\@addtoreset{equation}{subsection}
\renewcommand{\theequation}{\Alph{section}.\arabic{equation}}
\setcounter{section}{0}}
\makeatother

\ifx\href\asklfhas\newcommand{\href}[2]{#2}\fi

\def\pint#1 {- \!\!\!\!\!\!\!\! \,\int_{#1}}

\def\ni       {\noindent}

\def\semiket#1  { \, #1 \, \rangle \, }

\def\abs#1      {  \, \vert #1 \vert \,   }
\def\Im#1    { \, {\rm Im } \, #1  }
\def\Re#1    { \, {\rm Re}  \, #1  }

\def\binom#1#2 { \vecii{ {}_{#1} }{\raisebox{.5ex}{$ {}^{#2} $}} }
\def\sqbinom#1#2 { \Bigl(\begin{array}{c} {}_{#1}
                       \\ \raisebox{.5ex}{${}^{#2}$} \end{array}\Bigr)^2  }

\def\r12    {\frac{r_1}{r_2}}

\def\vecii#1#2      {  { #1 \choose #2 }  }
\def\veciii#1#2#3   {  \left(\begin{array}{c}#1\\#2\\#3\end{array}\right)  }
\def\matrixii#1#2#3#4            {  \Bigl( \begin{array}{cc}#1&#2\\#3&#4
                                       \end{array} \Bigr) }
\def\matrixiii#1#2#3#4#5#6#7#8#9 {  \left(\begin{array}{ccc}#1&#2&#3\\
                                     #4&#5&#6\\#7&#8&#9\end{array}\right)  }
\def\eqb         {  \begin{eqnarray}  }
\def\eqe           {  \end{eqnarray}  }

\def\csectionast#1    { \begin{center}
    {\large\bf #1  }   \end{center} \par \bigskip}
\def\titleandfile#1#2   {  \begin{center}{\large\bf #1}\end{center}
                            \par\begin{flushright} #2 \end{flushright} 
                            \par \begin{flushright} \today \end{flushright}}

\renewcommand{\thefootnote}{\fnsymbol{footnote}}

\begin{document}
%
\def\papertitlepage{\baselineskip 3.5ex \thispagestyle{empty}}
\def\preprinumber#1#2#3{\hfill \begin{minipage}{4.2cm}  #1
              \par\noindent #2
              \par\noindent #3
             \end{minipage}}
\renewcommand{\thefootnote}{\fnsymbol{footnote}}
%
%
\papertitlepage
\hfill UTHEP-548
\baselineskip 0.8cm
\vspace*{2.0cm}
\begin{center}
{\large\bf\mathversion{bold}
Microscopic formulation of the S-matrix in AdS/CFT}
\end{center}
\vskip 6ex
\baselineskip 1.0cm
\begin{center}
        { Kazuhiro ~Sakai\footnote[2]{\tt sakai@phys-h.keio.ac.jp}, } \\
 \vskip -1ex
    {\it Department of Physics, Keio University}
 \vskip -2ex   
    {\it Hiyoshi, Yokohama 223-8521, Japan} \\
 
 \vskip 2ex
     { Yuji  ~Satoh\footnote[3]{\tt ysatoh@het.ph.tsukuba.ac.jp}}  \\
 \vskip -1ex
    {\it Institute of Physics, University of Tsukuba} \\
 \vskip -2ex
   {\it Tsukuba, Ibaraki 305-8571, Japan}

\end{center}
\vskip 13ex
%
\baselineskip=3.5ex
\begin{center} {\bf Abstract} \end{center}

We develop the derivation we proposed in hep-th/0703177 of the
dressing phase of the S-matrix in the AdS/CFT correspondence in the
framework of the underlying bare integrable model. We elaborate the
configuration of the Bethe roots describing the physical vacuum, which
consists of a long Bethe string stretched along the imaginary axis and
stacks distributed along the real axis. We determine the distribution
of all Bethe roots in the thermodynamic limit. We then directly
compute the scattering phase of the fundamental excitations over the
physical vacuum and reproduce the BHL/BES dressing phase.

\vskip 2ex
%
%
%
%
%
\vspace*{\fill}
\ni
September 2007
\setcounter{page}{0}
\newpage
\renewcommand{\thefootnote}{\arabic{footnote}}
\setcounter{footnote}{0}
\setcounter{section}{0}
\baselineskip = 3.3ex
\pagestyle{plain}
\section{Introduction}

Integrability has been
providing us with new insights into
the duality between the ${\cal N}=4$ super Yang--Mills theory
and the superstring theory in $AdS_5\times S^5$.
After the discovery of integrability in the
one-loop super Yang--Mills theory
\cite{Minahan:2002ve,Beisert:2003yb}
and in the classical superstring theory\cite{Bena:2003wd},
a lot of progress has been made
toward the all-order/quantum integrability
in the full theory of
the planar AdS/CFT correspondence.
As a monumental result,
there has emerged a novel integrable model
\cite{Beisert:2005fw,Beisert:2005tm}
which is expected to
describe the spectrum of the infinitely long Yang--Mills operators
as well as that of the infinitely long quantum strings
\cite{Frolov:2006cc,Arutyunov:2006ak,Klose:2006zd},
at arbitrary values of the 't Hooft coupling constant $\lambda$.

The integrable model is characterized by the dispersion
relation\cite{Beisert:2004hm} and
the S-matrix\cite{Beisert:2005tm,Beisert:2006qh,Arutyunov:2006yd}
of the fundamental particles.
The system exhibits the centrally extended
$\alg{psu}(2|2)\oplus\alg{psu}(2|2)$ symmetry.
Remarkably, the symmetry completely
determines the dispersion relation and also
the S-matrix up to an overall scalar factor\cite{Beisert:2005tm}.
Given the scalar factor as a function of the momenta and the coupling,
one can systematically
study the spectrum of the system
by making use of powerful techniques
developed for conventional integrable models,
such as the Bethe ansatz.

The determination of the scalar factor,
or equivalently its principal part called the dressing phase,
was one of the main outstanding problems in this field.
The entire expression of the dressing phase
as the strong coupling expansion
was first constructed \cite{Beisert:2006ib} so that
it satisfies the crossing symmetry\cite{Janik:2006dc}
and includes the previously known
first two terms
\cite{Arutyunov:2004vx,Hernandez:2006tk,Arutyunov:2006iu,Freyhult:2006vr}
which reproduce the semi-classical string spectrum.
Subsequently, a systematic way of its determination
as the weak coupling expansion was presented
\cite{Beisert:2006ez}:
The problem can be rephrased
in terms of the cusp anomalous dimension\cite{Eden:2006rx}
where the transcendentality principle\cite{Kotikov:2002ab},
with some empirical rules, fully
determines the dressing phase up to an overall multiplicative constant.
The constant is readily singled out
by comparison with
either the perturbative computation\cite{Bern:2006ew}
or the strong coupling result\cite{Beisert:2006ib}.
Ultimately
the weak coupling result is identified with the strong coupling
one by a sort of analytic
continuation\cite{Beisert:2006ez,Kotikov:2006ts}
and is nicely expressed in a closed integral formula.
We call it the BHL/BES dressing phase
after the authors of the articles\cite{Beisert:2006ib,Beisert:2006ez}.
Its properties, such as the pole structure\cite{Dorey:2007xn}
as well as the strong coupling limit
\cite{Benna:2006nd,Alday:2007qf,Kostov:2007kx,Casteill:2007ct},
have been further studied.

Despite the success in the determination,
the clear understanding of the scalar factor was still lacking.
The above procedures do not explain
why the scalar factor should exhibit its particular structure.
It is also unsatisfactory that these procedures require some
model-specific computation of the string/gauge theory.
Although there are some interesting results explaining part of
its structure
\cite{Dorey:2007xn,Chen:2007vs,Gromov:2006cq,Gromov:2007cd},
one would desire a comprehensive explanation.

Let us recall here that
in the field of integrable models,
there are two well-known approaches
for the computation of the S-matrices:
One is called the factorized bootstrap program or
the phenomenological computation
\cite{Zamolodchikov:1978xm},
the other is called
the direct calculation,
the microscopic derivation or the Bethe ansatz technique
\cite{Korepin:1979qq,Faddeev:1981ft,Andrei:1983cb}.

The former approach
is to compute the S-matrices as an inverse problem.
In two-dimensional massive relativistic integrable models,
two-body S-matrices of the fundamental particles satisfy
the unitarity, the factorizability, and the crossing symmetry.
These conditions constrain the form of the S-matrices
up to the CDD ambiguity.
The ambiguity can be removed by some additional requirements,
such as the absence of the poles corresponding to unphysical particles.

The latter approach is to compute the S-matrices as a direct problem.
For example,
in the anti-ferromagnetic Heisenberg spin-chain
the physical vacuum is the anti-ferromagnetic state
rather than the ferromagnetic state.
The anti-ferromagnetic state
is realized as a nontrivial solution
of the bare Bethe ansatz equations
built over the ferromagnetic reference state.
In other words, the physical vacuum
is constructed by filling up the Dirac sea
over the bare vacuum.
The R-matrix describes the scattering of the magnons,
which are the fundamental excitations over the bare vacuum.
On the other hand,
the S-matrix appears as the scattering matrix of the spinons,
which are the fundamental excitations over the physical
vacuum.\footnote{
The R-matrix (the scattering matrix of magnons) is proportional
to the S-matrix and has a trivial scalar factor.
It also satisfies the unitarity, the factorizability,
but does not satisfy the crossing symmetry.
In the latter approach one does not assume the crossing symmetry
anywhere.
Instead, the S-matrix, which describes the physical scattering, 
becomes crossing-symmetric automatically, 
even though one starts from the crossing-non-invariant R-matrix.}

The above mentioned determination \cite{Beisert:2006ib}
of the scalar factor of the AdS/CFT S-matrix
basically followed the former bootstrap program.
It is natural to expect that
one could determine the scalar factor
alternatively by the latter direct computation.
The idea of such nontrivial structure of the physical vacuum
in the context of the AdS/CFT correspondence has been sometimes
considered \cite{Klose:2006dd}.
A concrete hint was observed in a computation of 
all-order anomalous dimensions\cite{Rej:2007vm}.
The authors of \cite{Rej:2007vm} derived
the integral equation describing the
all-order anomalous dimensions of field strength operators
and found that there appear integral kernels
very similar to those describing the scalar factor.
Such kernels are generated by the elimination
of density functions of Bethe roots at nested levels.
In our previous article \cite{Sakai:2007rk},
we demonstrated that
a certain configuration of the Bethe roots at nested levels
indeed generates the dressing phase in the
all-order Bethe equations.\footnote{
After the submission of our previous article\cite{Sakai:2007rk},
there appeared a similar computation
in the revised version of \cite{Rej:2007vm}.
While the mechanism of the generation of the dressing phase
is essentially the same,
their formulation looks conceptually different from ours.
For instance, there appear Bethe roots at the nested levels
twice as many kinds as ours.
}

In this article, we investigate in detail
this microscopic formulation of the AdS/CFT S-matrix.
After a brief review of the S-matrix and the all-order
Bethe equations in Section 2,
we present in Section 3 the whole configuration
of the Bethe roots describing the physical vacuum.
The configuration consists of a long Bethe string stretched
along the imaginary axis and stacks distributed along the real axis.
The former part corresponds to the configuration of a pulsating string,
while the latter is analogous to the vacuum configuration
of the Hubbard model in the attractive case.
We determine the density distribution of the stacks.
In Section 4,
we subsequently compute the density of stacks
in the presence of fundamental excitations.
Using this, we directly compute the S-matrix
as the two-body scattering matrix of the fundamental
excitations over the physical vacuum,
to find precisely the BHL/BES dressing phase 
\cite{Beisert:2006ib,Beisert:2006ez}.
Section 5 is devoted to a discussion.
The derivation of the effective momentum phase
of stacks is presented in Appendix A.

\section{S-matrix and nested Bethe ansatz equations}

Let us start our discussion with
an introduction of some notations.
The S-matrix is most concisely expressed with the help of
the following parametrizations
\begin{equation}\label{defofx}
x^\pm(u)=x(u\pm\tfrac{i}{2}),\quad
x(u)=\frac{u}{2}\left(1+\sqrt{1-4g^2/u^2}\right).
\end{equation}
Here $u$ is an analogue of the rapidity parameter
and 
\begin{equation}
g=\frac{\sqrt{\lambda}}{4\pi}\,
\end{equation}
is the normalized coupling constant.
In terms of these parameters,
the momentum $p$ of a fundamental particle
is expressed as
\begin{equation}
e^{ip}=\frac{x^+}{x^-}\,.
\end{equation}

The scattering matrix appearing in the context of
the AdS/CFT correspondence
exhibits the following tensor product structure
\begin{equation}\label{tensorSmatrix}
\hS(p_k,p_j)=S_0(p_k,p_j)^2[\hR(p_k,p_j)\otimes\hR(p_k,p_j)],
\end{equation}
where $\hR(p_k,p_j)$ is the $\alg{su}(2|2)$ invariant R-matrix
 of size $16\times 16$
and $S_0(p_k,p_j)^2$ is the overall scalar factor. 
The form of the R-matrix is completely determined
by the symmetry \cite{Beisert:2005tm, Beisert:2006qh}.
There are some variations of the canonical form of the R-matrix,
depending on the choice of the basis.
Here we adopt the string theory basis \cite{Arutyunov:2006yd}
so that the R-matrix satisfies the ordinary Yang--Baxter
algebra.

The overall scalar factor is conventionally
 expressed as 
\begin{equation}\label{S_0^2}
S_0(p_k,p_j)^2=
\frac{x^-_k-x^+_j}{x^+_k-x^-_j}\,
\frac{1-g^2/x^+_k\,x^-_j}{1-g^2/x^-_k\,x^+_j}\,
e^{2i\theta(u_k,u_j)},
\end{equation}
where $2\theta(u_k,u_j)$ is called
the dressing phase \cite{Arutyunov:2004vx}.
If we regard the S-matrix (\ref{tensorSmatrix})
as the scattering matrix of physical particles,
the dressing phase turns out to be a nontrivial function.
Its form was recently determined
\cite{Beisert:2006ib, Beisert:2006ez}.
Let us call it BHL/BES dressing phase.
It is expressed as
\begin{equation}\label{dressingphase}
2\theta_{\rm phys}(u_k,u_j)
=2ig^2
\int_{-\infty}^\infty dt e^{itu_k}e^{-\frac{|t|}{2}}
\int_{-\infty}^\infty dt' e^{it'u_j}e^{-\frac{|t'|}{2}}
\left(\hK_{\rm d}(2gt,2gt')-\hK_{\rm d}(2gt',2gt)\right),
\end{equation}
where the Fourier transform is a skew combination of
the dressing kernel
\begin{equation}\label{K_d}
\hK_{\rm d}(t,t')=8g^2\int_0^\infty
dt'' \hK_1(t,2gt'')\frac{t''}{e^{t''}-1}\hK_0(2gt'',t')\,.
\end{equation}
The constituent kernels are given by
\begin{equation}\label{K_0K_1}
\hK_0(t,t')
=\frac{tJ_1(t)J_0(t')-t'J_0(t)J_1(t')}{t^2-{t'}^2}\,,\quad
\hK_1(t,t')
=\frac{t'J_1(t)J_0(t')-tJ_0(t)J_1(t')}{t^2-{t'}^2}\,,
\end{equation}
where $J_n(t)$ are Bessel functions of the first kind.

The goal of the present article is to derive 
this BHL/BES dressing phase
in the context of the underlying bare integrable model.
In other words, we describe the system
starting from a bare vacuum
where the scattering matrix of the fundamental excitations
has the same structure as (\ref{tensorSmatrix})--(\ref{S_0^2})
but with the trivial dressing phase
\begin{equation}\label{trivialphase}
2\theta_{\rm bare}(u_k,u_j)=0.
\end{equation}
In the bare description,
the physical S-matrix can be computed as the scattering matrix of
the fundamental excitations over the Fermi surface.

Given the form of the $S$-matrix, one can derive a set of
Bethe ansatz equations.
Let us consider the system of $N$ particles
in a periodic one-dimensional box of length $L$.
We impose integrability of the system,
namely the condition that any multi-body scattering
is factorized into a product of two-body
scatterings
described by the above S-matrix.
For simplicity we consider the case of zero total momentum
\begin{equation}\label{zerototalmom}
P=\sum_{j=1}^Np_j=0.
\end{equation}
The consistency conditions for the periodicity
give rise to the Yang equations
\begin{equation}\label{Yangeqs}
e^{ip_kL}=\prod_{j\ne k}^N\hS(p_k,p_j).
\end{equation}
These matrix equations are diagonalized
with the help of the nested Bethe ansatz.
Combined with the momentum condition (\ref{zerototalmom}),
the nested Bethe ansatz equations
can be expressed as the asymptotic all-order
Bethe ansatz equations\cite{Beisert:2005fw}
\begin{equation}\label{unitytotalmom}
1=\prod_{j=1}^{K_4}\frac{x^+_{4,j}}{x^-_{4,j}}\,,
\end{equation}
\begin{eqnarray}
\label{BAE1}
1\Eqn{=}
\prod_{j=1}^{K_2}\frac{u_{1,k}-u_{2,j}+i/2}{u_{1,k}-u_{2,j}-i/2}
\prod_{j=1}^{K_4}\frac{1-g^2/x_{1,k}\,x^+_{4,j}}
                      {1-g^2/x_{1,k}\,x^-_{4,j}}\,,\\
\label{BAE2}
1\Eqn{=}
\prod_{j\ne k}^{K_2}\frac{u_{2,k}-u_{2,j}-i}{u_{2,k}-u_{2,j}+i}
\prod_{j=1}^{K_3}\frac{u_{2,k}-u_{3,j}+i/2}{u_{2,k}-u_{3,j}-i/2}
\prod_{j=1}^{K_1}\frac{u_{2,k}-u_{1,j}+i/2}{u_{2,k}-u_{1,j}-i/2}\,,\\
\label{BAE3}
1\Eqn{=}
\prod_{j=1}^{K_2}\frac{u_{3,k}-u_{2,j}+i/2}{u_{3,k}-u_{2,j}-i/2}
\prod_{j=1}^{K_4}\frac{x_{3,k}-x^+_{4,j}}{x_{3,k}-x^-_{4,j}}\,,\\
\label{BAE4}
\left(\frac{x^+_{4,k}}{x^-_{4,k}}\right)^J\Eqn{=}
\prod_{j\ne k}^{K_4}\frac{u_{4,k}-u_{4,j}+i}{u_{4,k}-u_{4,j}-i}
\,e^{2i\theta(u_{4,k},u_{4,j})}
\prod_{j=1}^{K_1}\frac{1-g^2/x^-_{4,k}\,x_{1,j}}
                      {1-g^2/x^+_{4,k}\,x_{1,j}}
\prod_{j=1}^{K_3}\frac{x^-_{4,k}-x_{3,j}}{x^+_{4,k}-x_{3,j}}
\no\\ &&\phantom{
\prod_{j\ne k}^{K_4}\frac{u_{4,k}-u_{4,j}+i}{u_{4,k}-u_{4,j}-i}
\,e^{2i\theta(u_{4,k},u_{4,j})}
}\hspace{-1.08em}\times
\prod_{j=1}^{K_7}\frac{1-g^2/x^-_{4,k}\,x_{7,j}}
                      {1-g^2/x^+_{4,k}\,x_{7,j}}
\prod_{j=1}^{K_5}\frac{x^-_{4,k}-x_{5,j}}{x^+_{4,k}-x_{5,j}}
\,,\quad\\
\label{BAE5}
1\Eqn{=}
\prod_{j=1}^{K_6}\frac{u_{5,k}-u_{6,j}+i/2}{u_{5,k}-u_{6,j}-i/2}
\prod_{j=1}^{K_4}\frac{x_{5,k}-x^+_{4,j}}{x_{5,k}-x^-_{4,j}}\,,\\
\label{BAE6}
1\Eqn{=}
\prod_{j\ne k}^{K_6}\frac{u_{6,k}-u_{6,j}-i}{u_{6,k}-u_{6,j}+i}
\prod_{j=1}^{K_5}\frac{u_{6,k}-u_{5,j}+i/2}{u_{6,k}-u_{5,j}-i/2}
\prod_{j=1}^{K_7}\frac{u_{6,k}-u_{7,j}+i/2}{u_{6,k}-u_{7,j}-i/2}\,,\\
\label{BAE7}
1\Eqn{=}
\prod_{j=1}^{K_6}\frac{u_{7,k}-u_{6,j}+i/2}{u_{7,k}-u_{6,j}-i/2}
\prod_{j=1}^{K_4}\frac{1-g^2/x_{7,k}\,x^+_{4,j}}
                      {1-g^2/x_{7,k}\,x^-_{4,j}}\,.
\end{eqnarray}
The length $L$ and the number of particle $N$
are interpreted as
\begin{equation}
L=J-K_4+\frac{1}{2}(-K_1+K_3+K_5-K_7),\quad N=K_4.
\end{equation}
We refer to
\cite{Beisert:2005tm,Beisert:2006qh,Martins:2007hb,deLeeuw:2007uf}
for the details of the derivation.

\section{Bethe root configuration of the physical vacuum}

In the bare description, physical states
are characterized by solutions of
the bare Bethe ansatz equations, that is,
the simultaneous equation (\ref{unitytotalmom})--(\ref{BAE7})
with the trivial dressing phase (\ref{trivialphase}).
In this section we present
a particular solution
that should express
the nontrivial physical vacuum state.

\subsection{General structure}

The configuration consists of the following occupation
numbers of bare Bethe roots\footnote{
The occupation numbers (\ref{occupation}) satisfy
the condition $K_2\le K_1+K_3\le K_4 \ge K_5+K_7\ge K_6$
required for the all-order Bethe ansatz equations.
(This condition follows from the consistency of
nested Bethe ansatz. See, e.g. \cite{Martins:2007hb}.)
On the other hand, they are outside the bound
$K_1\le K_2\le K_3\le K_4 \ge K_5\ge K_6\ge K_7$
required for the one-loop Bethe ansatz equations.
This means that the vacuum configuration
is characteristic of all-order Bethe ansatz equations and
becomes singular in the one-loop limit.}
\begin{equation}\label{occupation}
(K_1,\ldots,K_7)=(2M,M,0,2M,0,M,2M).
\end{equation}
For the vacuum state the configuration
of Bethe roots must be symmetric
with respect to the interchange of the two
$\alg{su}(2|2)$ sectors:
distribution of roots $u_{1,k},u_{2,k},u_{3,k}$ is just the same as
that of $u_{7,k},u_{6,k},u_{5,k}$, respectively.
Regarding this symmetry, we mostly omit to mention
the former copy of roots hereafter.

The vacuum has to be neutral
with respect to the pair of $\alg{su}(2|2)$ symmetries.
This restricts the relative numbers of the Bethe roots
to be $K_4=K_5+K_7=2K_6$. It can be understood as follows:
We restrict ourselves on one of the $\alg{su}(2|2)$'s.
Let $[n_1;n_2]$ denote the two $\alg{su}(2)$ charges,
by which we mean 
the Dynkin indices with respect to the bosonic subalgebra
$\alg{su}(2)\oplus\alg{su}(2)\subset \alg{su}(2|2)$.
A bosonic root $u_4$ creates a magnon with charges $[1;0]$.
Either of fermionic roots $u_5$ or $u_7$
converts the magnon charges $[1;0]$ to $[0;1]$.
A bosonic root $u_6$ flips the latter $\alg{su}(2)$ spin down,
namely it converts $[0;1]$ to $[0;-1]$.
Either $u_5$ or $u_7$ converts $[0;-1]$ to $[-1;0]$.
In other words, $u_5$ and $u_7$ have charges $[-1;1]$
while $u_6$ has charges $[0;-2]$.
It then follows that a state with general excitations
has charges $[K_4-K_5-K_7;K_5+K_7-2K_6]$.

The distribution among $K_5$ and $K_7$ is not determined
by the neutralness of the vacuum, since
the Bethe roots $u_{5}$ and $u_{7}$
originate in the same nested level of
diagonalization \cite{Beisert:2005tm}
and thus carry the same $\alg{su}(2|2)$ charges.
In fact, the Bethe roots $x_{7,k}$ are introduced
by the relabeling $x_{7}=g^2/x_{5}$ \cite{Beisert:2005fw}
in order to recover the seven sets of equations
(\ref{BAE1})--(\ref{BAE7})
out of five \cite{Beisert:2005tm, Martins:2007hb}.

Distinction of the roots $u_5$ and $u_7$ arises
in connection
with the $\alg{su}(2,2|4)$ one-loop Bethe equations.
For a general value of $u$,
the value of $x(u)$ has an ambiguity of
the square root branch (\ref{defofx}).
The branch of $x_5$ and $x_7$ are chosen
so that $x_5, x_7$ approach $u_5, u_7$, respectively,
in the one-loop limit $g\to 0$.
In other words, we relabel $x_{5}$ and $x_{7}$
through the relation $x_{7}=g^2/x_{5}$
so that all $x_5$'s and $x_7$'s
satisfy $|x_{5}|>g,|x_{7}|>g$.
The vacuum configuration has no $x_5$ root,
hence all the roots at the nested levels
decouple from $u_4$'s in the one-loop limit.
The set of occupation numbers (\ref{occupation})
are not allowed for the one-loop Bethe equations.
The vacuum configuration is
characteristic of the all-order Bethe equations.

When we derive the dressing phase, we send
both $J$ and $M$ to infinity. However,
for the purpose of studying the vacuum configuration,
it is convenient to take the limit $J\to\infty$ first
while keeping $M$ sufficiently large but finite.
We postpone taking the limit $M\to\infty$ until
we discuss excited states in the next section.

In what follows we will specify the whole configuration of
the bare Bethe roots.

\subsection{Configuration of the central Bethe roots}

In our previous article \cite{Sakai:2007rk},
we discussed mostly the configuration of
Bethe roots other than the $u_4$ roots, which 
is the most essential part of the derivation
of the dressing phase. Here, we further specify
the precise configuration of the $u_4$ roots
of the vacuum state.

The configuration of $u_4$'s
is extremely simple
when we see it on the $u$-plane.
For $J\to\infty$, it is given by
\begin{equation}\label{vacconfu4}
u_{4,k} = \tk i
\end{equation}
in terms of a shifted index $\tk=k-M-\tfrac{1}{2}$, which runs over
\begin{equation}
\tk=-M+\tfrac{1}{2},-M+\tfrac{3}{2},\ldots,M-\tfrac{3}{2},M-\tfrac{1}{2}.
\end{equation}
This configuration looks like nothing but
a conventional Bethe string of length $2M$.
In the present case, however,
it is not enough to specify only the values of $u_{4,k}$'s
because for each $x_{4,k}$ there is a choice of two branches
of the square root (\ref{defofx}). We choose them in such a way that
${\rm Im}\,x^+_{4,k}>0,\ {\rm Im}\,x^-_{4,k}<0$
for all roots. In other words, the vacuum configuration
is completely specified on the $x$-plane.
Explicitly, it is given by
\begin{equation}\label{vacconfx4}
x^\pm_{4,k}=\frac{i}{2}\left(\tk\pm\tfrac{1}{2}
  \pm\sqrt{(\tk\pm\tfrac{1}{2})^2+4g^2}\right).
\end{equation}
Note that the distribution (\ref{vacconfu4}) on the $u$-plane
is common to the magnon bound state \cite{Dorey:2006dq}.
For that state, however, the choice of branches is
$x^+_{4,k}=x^-_{4,k+1}$ for $k=1,\ldots,2M-1$
and ${\rm Im}\,x^-_{4,1}<0,\ {\rm Im}\,x^+_{4,2M}>0$,
which is different from (\ref{vacconfx4}).

Several comments are in order.
First, it is very natural that
the configuration is simple and, in particular,
does not have any continuous modulus parameter.
There is only one discrete parameter,
the number of $u_4$ roots $2M$,
which will be eventually sent to infinity.

Second, the vacuum configuration (\ref{vacconfu4}) for large $M$
is transparent when scattered with extra $u_4$ roots.
More precisely, when one scatters an extra root $u_4$
with the vacuum configuration (\ref{vacconfu4}),
it gains a scattering phase against each constituent
$u_{4,k}$. However, there occurs cancellation
and thus the total scattering phase is
\begin{equation}
\prod_{j=1}^{2M}\frac{u_4-u_{4,j}+i}{u_4-u_{4,j}-i}
=\frac{u_4+(M+\tfrac{1}{2})i}{u_4-(M+\tfrac{1}{2})i}\,
 \frac{u_4+(M-\tfrac{1}{2})i}{u_4-(M-\tfrac{1}{2})i}\,,
\end{equation}
which becomes trivial in the large $M$ limit.
This property is common to the magnon bound states
and is crucial later in the computation
of the dressing phase where we in fact add
extra $u_4$ roots to the vacuum.
On the other hand, 
in contrast to the case of the magnon bound state,
there occurs no cancellation
in the parts where $x_{4,k}$'s appear explicitly in the Bethe equations.
This is necessary
for having a sufficient number of stack solutions;
otherwise such a cancellation in (\ref{BAE7})
decreases the number of solutions of $u_{7,k}$
satisfying $|x_{7,k}|>g$ less than $2M$.

Third, the vacuum solution is characteristic of the 
all-order Bethe ansatz
equations: If we take the one-loop limit $g\to 0$,
the configuration (\ref{vacconfx4}) becomes singular,
which is in accord with the
fact that the dressing phase vanishes in this limit.
This is again in contrast to the magnon bound state,
which survives for $g\to 0$ with
each pair of $x^\pm_{4,k}$ 
approaching $u_{4,k}\pm i/2$.

Some readers might wonder whether the above Bethe string
with the present branch choice really exists,
though it solves the Bethe equations in the limit $J \to \infty$,  
and how to understand such a singular behavior in the one-loop limit.
To answer this question,
it would be instructive to consider the configuration
temporarily in the physical Bethe equations,
where we can make use of the correspondence with classical
strings.
The deviation due to the presence of the dressing phase
is within the error of the string hypothesis and is
negligible for $J\to\infty$.
Let us consider the thermodynamic limit 
$J\to\infty$,
keeping $g,M$ proportional to $J$,
and introduce a rescaled spectral parameter $\tx=x/g$.
On the $\tx$-plane
the imaginary roots (\ref{vacconfx4})
form two condensates 
$[-ib,-ib^{-1}],\ [ib^{-1},ib]$
with $b=M/2g+\sqrt{(M/2g)^2+1}$.
Because the configuration is symmetric
under the interchange $\tilde{x}\leftrightarrow 1/\tilde{x}$,
the corresponding classical string lives
in the $S^2\times\bbR$ sector \cite{Beisert:2004ag}.
There are few candidates for the solution
in the $S^2\times\bbR$ sector with only one modulus $b$.
We identify it as a pulsating string
\cite{Minahan:2002rc,deVega:1994yz,Kazakov:2004qf}
(see also \cite{Minahan:2006bd,Vicedo:2007rp}).
Pulsating string is an elliptic solution
and has a continuous elliptic modulus $k$ 
and a discrete winding number.
The winding number is read
from the density of the imaginary Bethe roots on the $u$-plane,
which is $1$ in this case.
Given the winding number,
the elliptic modulus $k$ is determined by $b$.
When we send $M$ to infinity, $b$ also goes to infinity
and $k$ approaches $0$.\footnote{
Although one can take $k$ arbitrarily small,
it cannot be strictly zero
as far as the winding number is nonzero.
The strictly rational case
$k=0$ corresponds to zero winding number,
which looks no longer a pulsating string
but rather a point-like string.}
Thus the configuration
(\ref{vacconfx4}) with large $M$ corresponds 
to the rational limit of the pulsating string.
It sweeps the $S^2$ at almost constant speed
with high frequency.

An unusual feature of this configuration is that
the condensates run across the unit circle on the $\tx$-plane.
Such a solution is precisely
an exception to the general correspondence between
classical strings and solutions of
the one-loop Bethe equations
\cite{Kazakov:2004qf,Beisert:2005bm,Beisert:2005di}.

Pulsating string solution has zero angular momentum
in the $S^2$. This is akin to the neutralness of the
anti-ferromagnetic state in a spin-chain.

\subsection{Formation of stacks}

In our vacuum configuration, the Bethe roots $u_6$ and $u_7$ form
stacks\cite{Sakai:2007rk}
\begin{equation}\label{stackrelation}
u_{7,2k-1}=u_{6,k}+\tfrac{i}{2},\quad
u_{7,2k}=u_{6,k}-\tfrac{i}{2},\quad
\mbox{for}\quad k=1,\ldots,M.
\end{equation}
Without knowing the bare Hamiltonian,
one cannot verify that this configuration
really corresponds to the ground state.
However,
most likely it does,
by analogy with the Hubbard model.
The Bethe equations (\ref{BAE6})--(\ref{BAE7}) resemble
very much the Lieb--Wu equations for the one-dimensional
Hubbard model.
The vacuum of the Hubbard model was well studied
\cite{Lieb:1968,Woynarovich:1983}.
In the attractive case, the vacuum
consists of precisely this kind of stacks\cite{Woynarovich:1983},
namely a kind of $k$--$\Lambda$ strings \cite{Takahashi:1972}.
Note that this kind of
stack also appears
in the description of the field strength operators
Tr ${\cal F}^L$ \cite{Rej:2007vm}.

Multiplying Bethe equations for $u_{7,2k-1}$
and for $u_{7,2k}$ together, one obtains the following set
of Bethe equations
\begin{equation}\label{BAE6center}
1=\prod_{j=1}^{2M}
\frac{1-g^2/x^+_{6,k}\,x^+_{4,j}}{1-g^2/x^+_{6,k}\,x^-_{4,j}}
\,\frac{1-g^2/x^-_{6,k}\,x^+_{4,j}}{1-g^2/x^-_{6,k}\,x^-_{4,j}}
\prod_{j\ne k}^{M}\frac{u_{6,k}-u_{6,j}+i}{u_{6,k}-u_{6,j}-i}\,.
\end{equation}
They can be viewed as effective Bethe equations
for $u_{6,k}$ denoting the centers of stacks.

\subsection{Distribution of stacks}

Given the configuration of $x^\pm_{4,k}$,
(\ref{BAE6center}) can be viewed as
Bethe equations for a single kind of Bethe roots with
a regular form of self-interaction:
\begin{equation}\label{stackBAE}
e^{i\Phi(u_{6,k})}=
  \prod_{j\ne k}^{M}\frac{u_{6,k}-u_{6,j}+i}{u_{6,k}-u_{6,j}-i}\,,
\end{equation}
where 
\begin{equation}
\Phi(u_{6,k})=\frac{1}{i}\sum_{j=1}^{2M}\ln
\frac{1-g^2/x^+_{6,k}\,x^-_{4,j}}{1-g^2/x^+_{6,k}\,x^+_{4,j}}
\,\frac{1-g^2/x^-_{6,k}\,x^-_{4,j}}{1-g^2/x^-_{6,k}\,x^+_{4,j}}\,
\end{equation}
is regarded as the virtual momentum phase.
For sufficiently large $M$,
one can evaluate this phase function by approximating sum by integral.
We relegate the detail of calculation to Appendix A.
If we take $M$ and $u$ sufficiently large
compared to the coupling constant $g$, 
the phase function approaches a reasonably simple form
\begin{equation}\label{phi}
\Phi(u)=
2M\left[2\arctan\frac{u}{M}
+\frac{u}{M}\ln\left(1+\frac{M^2}{u^2}\right)\right].
\end{equation}
An important property of the function $\Phi(u)$ is that
it is a monotonically increasing function.
This is clear from the form of its derivative
\begin{equation}
\Phi'(u)=2\ln\left(1+\frac{M^2}{u^2}\right).
\end{equation}
The Bethe equations (\ref{stackBAE}) are thus
analogous to those of the one-dimensional Bose gas with 
repulsive $\delta$-function interaction or
those of the $\alg{sl}(2,\mathbb{R})$ spin-chain.

The Bethe equations (\ref{stackBAE}) can be written
in the logarithmic form
\begin{equation}\label{stacklogBAE}
2\pi n_k=\Phi(u_k)+2\sum_{j\ne k}^M\arctan(u_k-u_j).
\end{equation}
We often abbreviate $u_{6,k}$ as $u_k$ hereafter.
The mode number $n_k$ associated with the root $u_k$
takes integer/half-integer value,
depending on $M$ is odd/even, respectively.
Since the r.h.s.\ is monotonically increasing
as a function of $u_k$, it follows that $n_k>n_j$ for $u_k>u_j$.
For the vacuum configuration,
we consider consecutive set of mode numbers.
One can always relabel the $u_k$ roots so that $u_k>u_j$ for $k>j$.
The mode numbers for the vacuum configuration are then given by
\begin{equation}\label{modenumbers}
n_k=-\tfrac{M-1}{2},-\tfrac{M-3}{2},\ldots,
  \tfrac{M-3}{2},\tfrac{M-1}{2}\quad
\mbox{for}\quad k=1,\ldots,M.
\end{equation}

In contrast to the case of the anti-ferromagnetic vacuum
of the Heisenberg chain, the present configuration
does not correspond to the maximal filling over the real axis.
In other words, the support of the distribution
of $u_k$'s is a finite interval.
One can see this as follows:
If the real axis were occupied by $u_1,\ldots,u_M$,
an extra real root with mode number $\tfrac{M+1}{2}$
would have to sit at $u=\infty$ \cite{Faddeev:1981ft}.
However, for $u_k=\infty$ the r.h.s.\ of (\ref{stacklogBAE})
would take $(3M-1)\pi$ and
thus $u_j<\infty$ for $2\pi n_j<(3M-1)\pi$,
which is contradictory to the last argument.

We are interested in the distribution of
$u_k$'s in the large $M$ limit.
From the form of the potential (\ref{phi}),
we see that the characteristic length of
the distribution of $u_k$'s is of order $M$.
Regarding this, let us expand the summand
of the interaction term in (\ref{stacklogBAE}) as
\begin{equation}
\arctan(u_k-u_j)=\frac{\pi}{2}{\rm sign}(u_k-u_j)
-\frac{1}{u_k-u_j}+{\cal O}\left(\frac{1}{(u_k-u_j)^3}\right)
\end{equation}
and evaluate the summation term by term. One finds that
the sum of the first term
precisely gives rise to the mode number
\begin{equation}
\frac{1}{2}\sum_{j\ne k}^M{\rm sign}(u_k-u_j)=n_k,
\end{equation}
while the sum of the lower order terms
after the second one becomes negligible.
The Bethe equations (\ref{stacklogBAE}) then reduce to
\begin{equation}
\Phi(u_k)=2\sum_{j\ne k}^M\frac{1}{u_k-u_j}\,.
\end{equation}
In the continuous limit,
one can replace the sum by the principal-value
integral and obtains
\begin{equation}\label{inteqrho}
\Phi(u)=2\vpint_{-B}^{B}\frac{\rho(v)dv}{u-v}\,,
\end{equation}
where we introduce the density function as
\begin{equation}
\rho(u)=\sum_{j=1}^M\delta(u-u_j).
\end{equation}
As we mentioned above the density has a finite support,
which is denoted by $[-B,B]$.

The integral equation (\ref{inteqrho}) can be solved by
the inverse Hilbert transformation
\begin{equation}\label{vacdensity}
\rho(u)=\frac{1}{2\pi^2}\vpint_{-B}^{B}
\sqrt{\frac{B^2-u^2}{B^2-v^2}}\,\frac{\Phi(v)dv}{v-u}\,.
\end{equation}
The endpoints $\pm B$ are determined by the normalization condition
\begin{equation}\label{rhonormalization}
\int_{-B}^{B}\rho(u)du=M.
\end{equation}
By evaluating these integral expressions,
one obtains the distribution of the stacks.

For our later purpose, let us estimate the order of $B$
with respect to $M$.
It is convenient to
rewrite the above equations in terms of the rescaled variables
\begin{equation}\label{rescaling}
u=M\tu,
\quad B=M\tB.
\end{equation}
We also introduce the normalized functions
\begin{equation}\label{nphi}
\tPhi(\tu)=
2\left[2\arctan\tu
+\tu\ln\left(1+\frac{1}{\tu^2}\right)\right],
\end{equation}
\begin{equation}\label{nrho}
\trho(\tu)=\frac{1}{M^2}\sum_{j=1}^M\delta(\tu-\tu_j),
\end{equation}
which are related to the original functions by
\begin{equation}
\Phi(u)=M\tPhi(\tu),\quad
\rho(u)=M\trho(\tu).
\end{equation}
In terms of these rescaled quantities,
the integral equation (\ref{inteqrho}) gives rise to
\begin{equation}\label{ninteqrho}
\tPhi(\tu)=2\vpint_{-\tB}^{\tB}\frac{\trho(\tv)d\tv}{\tu-\tv}\,.
\end{equation}
This equation is formally the same as (\ref{inteqrho}),
thus the solution is given by (\ref{vacdensity})
with all quantities replaced by the rescaled ones.
Note that $M$-dependence now enters only through
the endpoint value $\tB$, which is determined by
the normalization condition
\begin{equation}\label{nrhonorm}
\int_{-\tB}^{\tB} \trho(\tu)d\tu=\frac{1}{M}\,.
\end{equation}
It is clear that $\tB$ becomes small as one sends $M$ large.
This means that for large $M$,
$\trho(\tu)$ is determined
by the form of $\tPhi(\tu)$ only at small $\tu$.
Except for the very vicinity of the origin,
$\tPhi(\tu)$ at small $\tu$ roughly behaves as a linear function
\begin{equation}
\tPhi(\tu)\sim 4\tu\ln\frac{1}{\tB_0}\,,
\end{equation}
where $\tB_0\sim \tB$ is a typical scale.
With this approximation one can
analytically solve the integral equation
and obtains
\begin{equation}
\trho(\tu)\sim\frac{2}{\pi}\left(\ln\frac{1}{\tB_0}\right)
\sqrt{\tB^2-\tu^2}.
\end{equation}
The normalization condition (\ref{nrhonorm}) now reads
\begin{equation}
\tB^2\ln\frac{1}{\tB_0}\sim \frac{1}{M}\,.
\end{equation}
Ignoring the correction coming from the logarithm,
one finds that $\tB$ roughly scales with $M^{-1/2}$.
Thus the original $B$ roughly scales with $M^{1/2}$.

\section{Excited states and computation of the scattering phase}

In this section we consider excited states
and compute the two-body S-matrix of fundamental excitations.
By making use of the underlying symmetry,
the S-matrix can be written most generally
in the form of the spectral decomposition.
As the symmetry fixes the form of the projectors,
it is enough to determine
the eigenvalues in front of the projectors.
For example, the $\alg{su}(2)$ Zamolodchikovs' S-matrix
is constructed by computing the scattering phases
with respect to the triplet and the singlet.
In the present case, the centrally extended
$\alg{su}(2|2)$ algebra possesses a peculiar feature that
the tensor product of a pair of
4-dimensional atypical representations is irreducible
\cite{Beisert:2006qh}.
Therefore it is enough to compute only one scattering phase
of a pair of fundamental excitations
in a representative state.\footnote{
The computation presumes that the physical vacuum is a singlet.
In the last section we constructed the vacuum
as a neutral state under the pair of $\alg{su}(2|2)$ symmetries.
For the centrally extended algebra, however, the state
has to be neutral with respect to the central charges as well.
We define the action of the central charges
in our bare integrable model so that
the physical vacuum has zero central charges,
by shifting one of the central charges of the reference vacuum.}
This allows us to restrict our consideration
of excited states to those with only $u_4$ roots added.

\subsection{Fundamental excitations}

Let us consider excited states by adding extra $u_4$ roots
to the vacuum configuration.
We let $\cu_{4,k}$ denote the extra roots
and $\cK_4$ be their total number.
The occupation numbers read
\begin{equation}\label{multiu4occup}
(K_1,\ldots,K_7)=(2M,M,0,2M+\cK_4,0,M,2M).
\end{equation}
We keep the structure of the other roots unchanged,
namely the other $2M$ $u_4$'s constitute the Bethe string
with the branch choice (\ref{vacconfx4}), and 
the $u_6$'s and the $u_7$'s form
$M$ stacks with consecutive mode numbers.
In this subsection
let us determine the deviation
of the density of stacks for fixed $\cu_4$'s.

The effective Bethe equations for the centers of stacks read
\begin{equation}\label{stacklogBAEex}
2\pi n_k=\Phi_{\rm ex}(u_k)+2\sum_{j\ne k}^M\arctan(u_k-u_j),
\end{equation}
with mode numbers (\ref{modenumbers}).
The only difference from (\ref{stacklogBAE})
is the momentum phase
\begin{equation}
\Phi_{\rm ex}(u)=\Phi(u)+\varphi(u),
\end{equation}
where the modification part is given by
\begin{equation}
\varphi(u_{6,k})=
\frac{1}{i}\sum_{j=1}^{\cK_4}\ln
\frac{1-g^2/x^+_{6,k}\,\cx^-_{4,j}}{1-g^2/x^+_{6,k}\,\cx^+_{4,j}}
\,\frac{1-g^2/x^-_{6,k}\,\cx^-_{4,j}}{1-g^2/x^-_{6,k}\,\cx^+_{4,j}}\,
\end{equation}
with $\cx^\pm_{4,j} = x^\pm(\cu_{4,j})$.
By taking the derivative with respect to $u_k$,
(\ref{stacklogBAEex}) gives rise to
\begin{equation}\label{inteqrhoex}
2\pi\rho_{\rm ex}(u)=\Phi_{\rm ex}'(u)
+2\int_{-\Bex}^{\Bex}\frac{\rho_{\rm ex}(v)dv}{(u-v)^2+1}\,,
\end{equation}
where $\rho_{\rm ex}(u)$ is 
the density function of the centers of stacks
under the modified
momentum phase.
Apparently, one could derive the same integral equation for
the vacuum density $\rho(u)$ with the vacuum
phase $\Phi(u)$.
Subtracting it from (\ref{inteqrhoex}),
one obtains an integral equation
for the density deviation
$\sigma(u)=\rho_{\rm ex}(u)-\rho(u)$,
as follows
\begin{equation}
2\pi\sigma(u)=\varphi'(u)
+2\int_{-B}^B\frac{\sigma(v)dv}{(u-v)^2+1}
+2\left(\int_{-\Bex}^{-B}+\int_{B}^{\Bex}\right)
\frac{\rho_{\rm ex}(v)dv}{(u-v)^2+1}\,.
\end{equation}

We consider the large $M$ limit keeping the other parameters
$\cu_{4,k},\cK_4$ fixed. In this case the second integral is
negligible:
The fluctuation of the overall shape of $\Phi(u)$
should be suppressed at most within the change of $M$
to $M+\Delta M$ with $\Delta M={\cal O}(1)$.
Then
the deviation of $B\sim M^{1/2}$
is at most
$\Delta B\sim(M+\Delta M)^{1/2}-M^{1/2}
\sim M^{-1/2}\Delta M$.
On the other hand, we saw in the last section that
$\rho(u)\sim\ln(M/B_0)\sqrt{B^2-u^2}$.
Then the second integral is suppressed by
$\ln(M/B_0)B^{1/2}(\Delta B)^{3/2}\sim M^{-1/2}\ln M$,
which vanishes when $M$ is sent to infinity.
After all, in the limit $M\to\infty$ we obtain
\begin{equation}
2\pi\sigma(u)=\varphi'(u)
+2\int_{-\infty}^\infty\frac{\sigma(v)dv}{(u-v)^2+1}\,.
\end{equation}

This integral equation is solved
in the Fourier space \cite{Eden:2006rx,Rej:2007vm}.
Using the techniques in the Appendix D of \cite{Eden:2006rx},
one can derive the following formulas\footnote{
We assume $u,u'\in\bbR$.
The branch of logarithm should be chosen
appropriately.}
\begin{eqnarray}
\label{FourierHm}
\ln\left(1-g^2/x^\pm(u)\,x^\pm(u')\right)
\Eqn{=}\phantom{-}2g^2\!
\int_0^\infty dt e^{\pm iut}e^{-t/2}
\int_0^\infty dt' e^{\pm iu't'}e^{-t'/2}
{\hat H}_{\rm m}(2gt,2gt'),\\
\label{FourierKm}
\ln\left(1-g^2/x^\pm(u)\,x^\mp(u')\right)
\Eqn{=}-2g^2\!
\int_0^\infty dt e^{\pm iut}e^{-t/2}
\int_0^\infty dt' e^{\mp iu't'}e^{-t'/2}
{\hat K}_{\rm m}(2gt,2gt'),\hspace{2.5em}
\end{eqnarray}
where
the integral kernels are expressed in terms of Bessel functions
$J_n(t)$ by
\begin{equation}
\hH_{\rm m}(t,t')=\frac{J_1(t)J_0(t')+J_0(t)J_1(t')}{t+t'}\,,\quad
\hK_{\rm m}(t,t')=\frac{J_1(t)J_0(t')-J_0(t)J_1(t')}{t-t'}\,.
\end{equation}
With these formulas, one immediately obtains the solution
in the Fourier space
\begin{equation}\label{sigmaFT}
\sigma(u)=\frac{1}{2\pi}\int_{-\infty}^\infty
dt e^{itu}\hat\sigma(t),
\end{equation}
where
\begin{equation}\label{sigmahat}
\hat\sigma(\pm t)=
-\frac{g^2t}{\sinh\frac{t}{2}}
\sum_{j=1}^{\cK_4}\int_0^\infty dt'e^{-t'/2}
\left(e^{\pm it'\cu_{4,j}}\hH_{\rm m}(2gt,2gt')
     +e^{\mp it'\cu_{4,j}}\hK_{\rm m}(2gt,2gt')\right),
\end{equation}
for $t>0$.

\subsection{The dressing phase}

We are now in a position to compute
the dressing phase.
The vacuum configuration in the bare description
corresponds to the empty state in the physical description.
The physical fundamental excitations are described by
adding extra roots $\cu_{4,k}$ in the bare description.
The bare configuration we studied in the last subsection
corresponds to the system of $\cK_4$ excitations.

In the physical description, the scattering phase of the two
fundamental excitations is simply given by
\begin{equation}\label{physshift}
\phi_{12}=\frac{1}{i}
\ln\frac{\cu_{4,1}-\cu_{4,2}+i}{\cu_{4,1}-\cu_{4,2}-i}
+2\theta_{\rm phys}(\cu_{4,1},\cu_{4,2}).
\end{equation}
In the bare description, the same scattering phase is
expressed by the difference of two phases
\begin{equation}\label{bareshift}
\phi_{12}=\delta_{12}(\cu_{4,1})-\delta_{1}(\cu_{4,1}).
\end{equation}
Here $\delta_{12}$ is the total phase which
the first excitation gains when moving around the chain
in the presence of the second excitation.
$\delta_1$ is measured in the same way
but in the absence of the second excitation \cite{Korepin:1979qq}.

The total phase is the phase of the transfer matrix eigenvalue
and thus can be read from the r.h.s.\ of
the central Bethe equations (\ref{BAE4}).
By substituting (\ref{stackrelation})
(and the corresponding relations for $u_{1,k},u_{2,k}$), they read
\begin{eqnarray}\label{BAE4bis}
\left(\frac{\cx^+_{4,k}}{\cx^-_{4,k}}\right)^J
\Eqn{=}
\prod_{j\ne k}^{\cK_4}
\frac{\cu_{4,k}-\cu_{4,j}+i}{\cu_{4,k}-\cu_{4,j}-i}
\prod_{j=1}^{2M}\frac{\cu_{4,k}-u_{4,j}+i}{\cu_{4,k}-u_{4,j}-i}
\no\\
&&\times
\prod_{j=1}^{M}\frac{1-g^2/\cx^-_{4,k}\,x^+_{2,j}}
                    {1-g^2/\cx^+_{4,k}\,x^+_{2,j}}
\prod_{j=1}^{M}\frac{1-g^2/\cx^-_{4,k}\,x^-_{2,j}}
                    {1-g^2/\cx^+_{4,k}\,x^-_{2,j}}
\no\\
&&\times
\prod_{j=1}^{M}\frac{1-g^2/\cx^-_{4,k}\,x^+_{6,j}}
                    {1-g^2/\cx^+_{4,k}\,x^+_{6,j}}
\prod_{j=1}^{M}\frac{1-g^2/\cx^-_{4,k}\,x^-_{6,j}}
                    {1-g^2/\cx^+_{4,k}\,x^-_{6,j}}
\,.
\end{eqnarray}
The total phase is then expressed as
\begin{eqnarray}
\delta_{1\ldots\cK_4}(\cu_{4,k})
\Eqn{=}
\frac{1}{i}\sum_{j\ne k}^{\cK_4}
\ln\frac{\cu_{4,k}-\cu_{4,j}+i}{\cu_{4,k}-\cu_{4,j}-i}
+\frac{1}{i}\sum_{j=1}^{2M}
\ln\frac{\cu_{4,k}-u_{4,j}+i}{\cu_{4,k}-u_{4,j}-i}\no\\
&&+\frac{2}{i}\int_{-\infty}^\infty
\ln\left[
\frac{1-g^2/\cx^-_{4,1}\,x^+(u)}{1-g^2/\cx^+_{4,1}\,x^+(u)}\,
\frac{1-g^2/\cx^-_{4,1}\,x^-(u)}{1-g^2/\cx^+_{4,1}\,x^-(u)}\right]
\rho_{\rm ex}(u) du.\quad
\end{eqnarray}
Therefore (\ref{bareshift}) gives rise to
\begin{eqnarray}
\phi_{12}
\Eqn{=}
\frac{1}{i}\ln\frac{\cu_{4,1}-\cu_{4,2}+i}{\cu_{4,1}-\cu_{4,2}-i}\no\\
&&+\frac{2}{i}\int_{-\infty}^\infty
\ln\left[
\frac{1-g^2/\cx^-_{4,1}\,x^+(u)}{1-g^2/\cx^+_{4,1}\,x^+(u)}\,
\frac{1-g^2/\cx^-_{4,1}\,x^-(u)}{1-g^2/\cx^+_{4,1}\,x^-(u)}\right]
\left(\sigma_{12}(u)-\sigma_{1}(u)\phantom{\rule{0ex}{2ex}}\right)
du,\quad
\end{eqnarray}
where $\sigma_{12}(u),\sigma_1(u)$ are given by
(\ref{sigmaFT})--(\ref{sigmahat}) with $\cK_4=2,1$, respectively.
The first term is the bare scattering phase
which the first excitation directly feels against the second one.
The second term comes from the scattering of the first
excitation against the stacks whose
density deviation $\sigma_{12}-\sigma_{1}$
encodes the back reaction from the second excitation.
From the comparison with (\ref{physshift}), we see that
the second term plays precisely the role of the dressing phase.
By using the formulas (\ref{FourierHm})--(\ref{FourierKm}) again,
the second term can be expressed concisely in the Fourier space
\begin{eqnarray}\label{dressingphase2}
&&\hspace{-2em}
2\theta_{\rm phys}(\cu_{4,1},\cu_{4,2})\no\\[1ex]
\Eqn{=}
2g^2\!
\int_{-\infty}^\infty dt e^{it\cu_{4,1}}e^{-\frac{|t|}{2}}
\int_{-\infty}^\infty dt' e^{it'\cu_{4,2}}e^{-\frac{|t'|}{2}}
\left(\hat{\cal K}(2gt,2gt')-\hat{\cal K}(2gt',2gt)\right),
\hspace{1em}
\end{eqnarray}
where
\begin{equation}
\hat{\cal K}(2gt,2gt')=4g^2\int_0^\infty
dt'' \hK_{\rm m}(2gt,2gt'')
\frac{it''}{e^{t''}-1}\hH_{\rm m}(2gt'',2gt')\,.
\end{equation}
This precisely agrees with
the BHL/BES dressing phase (\ref{dressingphase}).

\section{Discussion}

We have computed the two-body scattering
phase of the fundamental excitations over the physical vacuum,
which precisely agrees with the BHL/BES dressing phase.
By taking account of the centrally extended
$\alg{psu}(2|2)\oplus\alg{psu}(2|2)$ symmetry,
this suffices to determine the whole $256\times 256$
components of the S-matrix.
From this S-matrix, one can construct the Yang equations
and derive the complete set of physical Bethe ansatz equations,
as explained in Section 2.
Altogether, our formulation proposes a derivation
of the asymptotic all-order Bethe ansatz equations
with the BHL/BES dressing phase,
purely based on the symmetry and the integrability.

In the last section 
we have considered particular excited states
that consist of only $\cK_4$ excitations.
For these states the correspondence between the bare
description and the physical one may be trivial in the sense that
each extra bare root represents a physical root.
However, the correspondence is not so simple in general:
When the occupation numbers of the physical roots
are still equal to that of the extra bare roots,
the values of the roots could differ.
In general, addition of extra Bethe roots at nested levels
partly breaks the structure of the stacks. A single physical
root sometimes corresponds to a complex of bare roots and holes.
It would be interesting to clarify the correspondence.

For the moment we do not know whether
a Yang--Mills operator of finite length
can be directly realized in the bare description.
A possibility is that a physical operator of length $\cL$
could be expressed by a state
in the chain of length $L+\cL$
while the vacuum is defined in the chain of length $L$.
Of course both $L$ and $L+\cL$ have to be sent to infinity,
but still the difference would make sense.

We have determined our vacuum configuration
as the simplest consistent solution that generates
the BHL/BES dressing phase.
However, ultimately we wish to derive it
as the ground state of a certain Hamiltonian.
The Hamiltonian has to be expressed in a form
compatible with the bare description,
preferably in terms of the $\alg{su}(2|2)$ R-matrix.
It may be derived from the gauge-fixed light-cone
Hamiltonian for the Green--Schwarz superstring theory
in $AdS_5\times S^5$ \cite{Frolov:2006cc},
which exhibits the invariance
under the centrally extended
$\alg{psu}(2|2)\oplus\alg{psu}(2|2)$ symmetry
when the worldsheet is decompactified \cite{Arutyunov:2006ak}.

Our microscopic formulation will be of fundamental use
in various directions under the latest investigation,
for example, the boundary S-matrix\cite{Hofman:2007xp},
the wrapping interactions\cite{Ambjorn:2005wa,Janik:2007wt}
and the Baxter equations\cite{Kazakov:2007fy,Belitsky:2007cc}.
We hope to report the progress in these topics elsewhere.


%
\begin{center}
  {\bf Acknowledgments}
\end{center}

We are most grateful to D.~Serban and K.~Zarembo
for illuminating comments and discussions. K.S. would like to thank
N.~Dorey, N.~Gromov, V.~Kazakov, K.~Okamura,
R.~Suzuki, Y.~Takayama, P.~Vieira for comments and discussions
whereas Y.S. would like to thank N. Ishibashi for useful conversations.
K.S. also thanks the organizers of the 12th Claude Itzykson Meeting
held at the Service de Physique Th\'eorique, CEA Saclay
and at the Ecole Normale Sup\'erieure for their kind hospitality.

\appendix
\section{Effective momentum phase\label{App:EffPot}}

The centers of stacks $u_{6,k}$ obey
the following effective Bethe equations
\begin{equation}
e^{i\Phi(u_{6,k})}=
\prod_{j\ne k}^{M}\frac{u_{6,k}-u_{6,j}+i}{u_{6,k}-u_{6,j}-i}\,,
\end{equation}
where
\begin{eqnarray}
&&\Phi(u_{6,k})=\Phi^+(u_{6,k})+\Phi^-(u_{6,k}),\\[1ex]
&&\hspace{3em}
\Phi^\pm(u_{6,k})=\frac{1}{i}\sum_{j=1}^{2M}\ln
\frac{1-g^2/x^\pm_{6,k}\,x^-_{4,j}}{1-g^2/x^\pm_{6,k}\,x^+_{4,j}}\,.
\end{eqnarray}
Note that
\begin{equation}
g^2/x^\pm_{4,j}=\frac{i}{2}\Bigl(\tj\pm\tfrac{1}{2}
 \mp\sqrt{(\tj\pm\tfrac{1}{2})^2+4g^2}\Bigr),
\end{equation}
where the index $\tj$ is defined as
\begin{eqnarray}
\tj\Eqn{=}j-M-\tfrac{1}{2}\no\\
\Eqn{=}-M+\tfrac{1}{2},-M+\tfrac{3}{2},
  \ldots,M-\tfrac{3}{2},M-\tfrac{1}{2}.
\end{eqnarray}
Let us evaluate the function $\Phi^+(u)$ in the large $M$ limit:
\begin{eqnarray}
\Phi^+(u)\Eqn{=}\frac{1}{i}\sum_{j=1}^{2M}\ln
\frac{1-g^2/x^-_{4,j}\,x^+(u)}{1-g^2/x^+_{4,j}\,x^+(u)}\no\\
\Eqn{=}\frac{1}{i}\sum_{\tj=-M+1/2}^{M-1/2}
  \ln\frac{\sqrt{(\tj-\tfrac{1}{2})^2+4g^2}
                +(\tj-\tfrac{1}{2})+2ix^+(u)}
          {\sqrt{(\tj+\tfrac{1}{2})^2+4g^2}
                -(\tj+\tfrac{1}{2})-2ix^+(u)}\no\\
\label{sumfinal}
\Eqn{=}\frac{1}{i}\sum_{\tj=-M+1/2}^{M-1/2}
  \ln\frac{\sqrt{t_\tj^2+a^2}-t_\tj+b}{\sqrt{t_\tj^2+a^2}-t_\tj-b}\,,
\end{eqnarray}
where
\begin{equation}
t_\tj=\frac{\tj+\tfrac{1}{2}}{M},\quad a=\frac{2g}{M},
\quad b=\frac{2ix^+(u)}{M}\,.
\end{equation}
In the large $M$ limit, 
one can approximate the sum in (\ref{sumfinal}) by the integral
\begin{equation}
\Phi^+(u)
=\frac{M}{i}\int_{-1}^1
dt\ln\frac{\sqrt{t^2+a^2}-t+b}{\sqrt{t^2+a^2}-t-b}\,.
\end{equation}
This integral can be performed by the change of variable
$s=\sqrt{t^2+a^2}-t$. In fact,
\begin{eqnarray}
&&\hspace{-3em}\int_{-1}^1 dt \ln(\sqrt{t^2+a^2}-t+b)\no\\
\Eqn{=}\int_{\sqrt{1+a^2}+1}^{\sqrt{1+a^2}-1}
  ds\left(-\frac{s^2+a^2}{2s^2}\right)\ln(s+b)\no\\
\Eqn{=}\frac{1}{2}\left[
\Bigl(\frac{a^2}{s}-s\Bigr)\ln(s+b)-b\ln(s+b)+\frac{a^2}{b}
\ln\Bigl(1+\frac{b}{s}\Bigr)+s+b
\right]_{s=\sqrt{1+a^2}+1}^{s=\sqrt{1+a^2}-1}.
\end{eqnarray}
Using this, one obtains
\begin{eqnarray}
\Phi^+(u)\Eqn{=}
\frac{M}{2}\left[
2i\ln\frac{(\lambda+i\sqrt{1+a^2})^2+1}{(\lambda-i\sqrt{1+a^2})^2+1}
\right.\no\\ &&\hspace{3em}\left.
+\lambda\ln\frac{\lambda^2+(\sqrt{1+a^2}+1)^2}
                {\lambda^2+(\sqrt{1+a^2}-1)^2}
-\frac{a^2}{\lambda}\ln\frac{a^4/\lambda^2+(\sqrt{1+a^2}+1)^2}
                      {a^4/\lambda^2+(\sqrt{1+a^2}-1)^2}
\right],\qquad
\end{eqnarray}
where
\begin{equation}
\lambda=-ib=\frac{2x^+(u)}{M},\quad a=\frac{2g}{M}\,.
\end{equation}
$\Phi^-(u)$ takes the same form with $\lambda=2x^-(u)/M$.

Let us consider the case
where the coupling constant $g$ is finite.
As we take $M$ and $u$ sufficiently large, we see that
\begin{equation}
\lambda\approx\frac{2u}{M},\quad a\approx 0.
\end{equation}
As a result, the phase function reduces to a reasonably simple form
\begin{equation}
\Phi(u)=
2M\left[2\arctan\frac{u}{M}
+\frac{u}{M}\ln\left(1+\frac{M^2}{u^2}\right)\right] - 2 \pi M.
\end{equation}
In the above computation we implicitly chose the branch of logarithm
so that $\Phi(u)=0$ at $u=+\infty$.
In the main text we drop the constant $-2\pi M$,
which corresponds to the choice of the branch
where $\Phi(u)=0$ at $u=0$.

%
%
\def\thebibliography#1{\list
 {[\arabic{enumi}]}{\settowidth\labelwidth{[#1]}\leftmargin\labelwidth
  \advance\leftmargin\labelsep
  \usecounter{enumi}}
  \def\newblock{\hskip .11em plus .33em minus .07em}
  \sloppy\clubpenalty4000\widowpenalty4000
  \sfcode`\.=1000\relax}
 \let\endthebibliography=\endlist
%
%
\vspace{3ex}
\begin{center}
 {\bf References}
\end{center}
\par \vspace*{-2ex}

%

%


\begin{thebibliography}{999}
\parskip=-2.5pt
%

  
\bibitem{Minahan:2002ve}
  J.~A.~Minahan and K.~Zarembo,
  JHEP {\bf 0303} (2003) 013
  [arXiv:hep-th/0212208].

\bibitem{Beisert:2003yb}
  N.~Beisert and M.~Staudacher,
  Nucl.\ Phys.\  B {\bf 670} (2003) 439
  [arXiv:hep-th/0307042].

\bibitem{Bena:2003wd}
  I.~Bena, J.~Polchinski and R.~Roiban,
  Phys.\ Rev.\  D {\bf 69} (2004) 046002
  [arXiv:hep-th/0305116].

\bibitem{Beisert:2005fw}
  N.~Beisert and M.~Staudacher,
  Nucl.\ Phys.\  B {\bf 727} (2005) 1
  [arXiv:hep-th/0504190].

\bibitem{Beisert:2005tm}
  N.~Beisert,
  arXiv:hep-th/0511082.

\bibitem{Frolov:2006cc}
  S.~Frolov, J.~Plefka and M.~Zamaklar,
  J.\ Phys.\ A  {\bf 39} (2006) 13037
  [arXiv:hep-th/0603008].

\bibitem{Arutyunov:2006ak}
  G.~Arutyunov, S.~Frolov, J.~Plefka and M.~Zamaklar,
  J.\ Phys.\ A  {\bf 40} (2007) 3583
  [arXiv:hep-th/0609157].

\bibitem{Klose:2006zd}
  T.~Klose, T.~McLoughlin, R.~Roiban and K.~Zarembo,
  JHEP {\bf 0703} (2007) 094
  [arXiv:hep-th/0611169].

\bibitem{Beisert:2004hm}
  N.~Beisert, V.~Dippel and M.~Staudacher,
  JHEP {\bf 0407} (2004) 075
  [arXiv:hep-th/0405001].
 
\bibitem{Beisert:2006qh}
  N.~Beisert,
  J.\ Stat.\ Mech.\  {\bf 0701} (2007) P017
  [arXiv:nlin.si/0610017].

\bibitem{Arutyunov:2006yd}
  G.~Arutyunov, S.~Frolov and M.~Zamaklar,
  JHEP {\bf 0704} (2007) 002
  [arXiv:hep-th/0612229].

\bibitem{Beisert:2006ib}
  N.~Beisert, R.~Hern\'andez and E.~L\'opez,
  JHEP {\bf 0611} (2006) 070
  [arXiv:hep-th/0609044].

\bibitem{Janik:2006dc}
  R.~A.~Janik,
  Phys.\ Rev.\  D {\bf 73} (2006) 086006
  [arXiv:hep-th/0603038].
    
\bibitem{Arutyunov:2004vx}
  G.~Arutyunov, S.~Frolov and M.~Staudacher,
  JHEP {\bf 0410} (2004) 016
  [arXiv:hep-th/0406256].

\bibitem{Hernandez:2006tk}
  R.~Hern\'andez and E.~L\'opez,
  JHEP {\bf 0607} (2006) 004
  [arXiv:hep-th/0603204].

\bibitem{Arutyunov:2006iu}
  G.~Arutyunov and S.~Frolov,
  Phys.\ Lett.\  B {\bf 639} (2006) 378
  [arXiv:hep-th/0604043].

\bibitem{Freyhult:2006vr}
  L.~Freyhult and C.~Kristjansen,
  Phys.\ Lett.\  B {\bf 638} (2006) 258
  [arXiv:hep-th/0604069].

\bibitem{Beisert:2006ez}
  N.~Beisert, B.~Eden and M.~Staudacher,
  J.\ Stat.\ Mech.\  {\bf 0701} (2007) P021
  [arXiv:hep-th/0610251].

\bibitem{Eden:2006rx}
  B.~Eden and M.~Staudacher,
  J.\ Stat.\ Mech.\  {\bf 0611} (2006) P014
  [arXiv:hep-th/0603157].

\bibitem{Kotikov:2002ab}
  A.~V.~Kotikov and L.~N.~Lipatov,
  Nucl.\ Phys.\  B {\bf 661} (2003) 19
  [Erratum-ibid.\  B {\bf 685} (2004) 405]
  [arXiv:hep-ph/0208220].

\bibitem{Bern:2006ew}
  Z.~Bern, M.~Czakon, L.~J.~Dixon, D.~A.~Kosower and V.~A.~Smirnov,
  Phys.\ Rev.\  D {\bf 75} (2007) 085010
  [arXiv:hep-th/0610248].

\bibitem{Kotikov:2006ts}
  A.~V.~Kotikov and L.~N.~Lipatov,
  Nucl.\ Phys.\  B {\bf 769} (2007) 217
  [arXiv:hep-th/0611204].

\bibitem{Dorey:2007xn}
  N.~Dorey, D.~M.~Hofman and J.~Maldacena,
  Phys.\ Rev.\  D {\bf 76} (2007) 025011
  [arXiv:hep-th/0703104].

\bibitem{Benna:2006nd}
  M.~K.~Benna, S.~Benvenuti, I.~R.~Klebanov and A.~Scardicchio,
  Phys.\ Rev.\ Lett.\  {\bf 98} (2007) 131603
  [arXiv:hep-th/0611135].

\bibitem{Alday:2007qf}
  L.~F.~Alday, G.~Arutyunov, M.~K.~Benna, B.~Eden and I.~R.~Klebanov,
  JHEP {\bf 0704} (2007) 082
  [arXiv:hep-th/0702028].

\bibitem{Kostov:2007kx}
  I.~Kostov, D.~Serban and D.~Volin,
  Nucl.\ Phys.\  B {\bf 789} (2008) 413
  [arXiv:hep-th/0703031].

\bibitem{Casteill:2007ct}
  P.~Y.~Casteill and C.~Kristjansen,
  Nucl.\ Phys.\  B {\bf 785} (2007) 1
  [arXiv:0705.0890 [hep-th]].

\bibitem{Chen:2007vs}
  H.~Y.~Chen, N.~Dorey and R.~F.~Lima Matos,
  JHEP {\bf 0709} (2007) 106
  [arXiv:0707.0668 [hep-th]].

\bibitem{Gromov:2006cq}
  N.~Gromov and V.~Kazakov,
  Nucl.\ Phys.\  B {\bf 780} (2007) 143
  [arXiv:hep-th/0605026].

\bibitem{Gromov:2007cd}
  N.~Gromov and P.~Vieira,
  Nucl.\ Phys.\  B {\bf 790} (2008) 72
  [arXiv:hep-th/0703266].

\bibitem{Zamolodchikov:1978xm}
  A.~B.~Zamolodchikov and A.~B.~Zamolodchikov,
  Annals Phys.\  {\bf 120} (1979) 253.

\bibitem{Korepin:1979qq}
  V.~E.~Korepin,
  Theor.\ Math.\ Phys.\  {\bf 41} (1979) 953
  [Teor.\ Mat.\ Fiz.\  {\bf 41} (1979) 169].

\bibitem{Faddeev:1981ft}
  L.~D.~Faddeev and L.~A.~Takhtajan,
  J.\ Sov.\ Math.\  {\bf 24} (1984) 241
  [Zap.\ Nauchn.\ Semin.\  {\bf 109} (1981) 134].
    
\bibitem{Andrei:1983cb}
  N.~Andrei and C.~Destri,
  Nucl.\ Phys.\  B {\bf 231} (1984) 445.
  
\bibitem{Klose:2006dd}
  T.~Klose and K.~Zarembo,
  J.\ Stat.\ Mech.\  {\bf 0605} (2006) P006
  [arXiv:hep-th/0603039].

\bibitem{Rej:2007vm}
  A.~Rej, M.~Staudacher and S.~Zieme,
  J.\ Stat.\ Mech.\  {\bf 0708} (2007) P08006
  [arXiv:hep-th/0702151].

\bibitem{Sakai:2007rk}
  K.~Sakai and Y.~Satoh,
  arXiv:hep-th/0703177.

\bibitem{Martins:2007hb}
  M.~J.~Martins and C.~S.~Melo,
  Nucl.\ Phys.\  B {\bf 785} (2007) 246
  [arXiv:hep-th/0703086].

\bibitem{deLeeuw:2007uf}
  M.~de Leeuw,
  arXiv:0705.2369 [hep-th].

\bibitem{Dorey:2006dq}
  N.~Dorey,
  J.\ Phys.\ A  {\bf 39} (2006) 13119
  [arXiv:hep-th/0604175].

\bibitem{Beisert:2004ag}
  N.~Beisert, V.~A.~Kazakov and K.~Sakai,
  Commun.\ Math.\ Phys.\  {\bf 263} (2006) 611
  [arXiv:hep-th/0410253].

\bibitem{Kazakov:2004qf}
  V.~A.~Kazakov, A.~Marshakov, J.~A.~Minahan and K.~Zarembo,
  JHEP {\bf 0405} (2004) 024
  [arXiv:hep-th/0402207].

\bibitem{deVega:1994yz}
  H.~J.~de Vega, A.~L.~Larsen and N.~G.~Sanchez,
  Phys.\ Rev.\  D {\bf 51} (1995) 6917
  [arXiv:hep-th/9410219].

\bibitem{Minahan:2002rc}
  J.~A.~Minahan,
  Nucl.\ Phys.\  B {\bf 648} (2003) 203
  [arXiv:hep-th/0209047].

\bibitem{Minahan:2006bd}
  J.~A.~Minahan, A.~Tirziu and A.~A.~Tseytlin,
  JHEP {\bf 0608} (2006) 049
  [arXiv:hep-th/0606145].

\bibitem{Vicedo:2007rp}
  B.~Vicedo,
  arXiv:hep-th/0703180.

\bibitem{Beisert:2005bm}
  N.~Beisert, V.~A.~Kazakov, K.~Sakai and K.~Zarembo,
  Commun.\ Math.\ Phys.\  {\bf 263} (2006) 659
  [arXiv:hep-th/0502226].

\bibitem{Beisert:2005di}
  N.~Beisert, V.~A.~Kazakov, K.~Sakai and K.~Zarembo,
  JHEP {\bf 0507} (2005) 030
  [arXiv:hep-th/0503200].

\bibitem{Lieb:1968}
  E.~H.~Lieb and F.~Y.~Wu,
  Phys.\ Rev.\ Lett.\  {\bf 20} (1968) 1445.

\bibitem{Woynarovich:1983}
  F.~Woynarovich,
  J.\ Phys.\ C: Solid State Phys.\ {\bf 16} (1983) 6593.

\bibitem{Takahashi:1972}
  M.~Takahashi,
  Prog.\ Theor.\ Phys.\ {\bf 47} (1972) 69.

\bibitem{Hofman:2007xp}
  D.~M.~Hofman and J.~M.~Maldacena,
  arXiv:0708.2272 [hep-th].

\bibitem{Ambjorn:2005wa}
  J.~Ambj{\o}rn, R.~A.~Janik and C.~Kristjansen,
  Nucl.\ Phys.\  B {\bf 736} (2006) 288
  [arXiv:hep-th/0510171].

\bibitem{Janik:2007wt}
  R.~A.~Janik and T.~{\L}ukowski,
  arXiv:0708.2208 [hep-th].

\bibitem{Kazakov:2007fy}
  V.~Kazakov, A.~Sorin and A.~Zabrodin,
  Nucl.\ Phys.\  B {\bf 790} (2008) 345
  [arXiv:hep-th/0703147].

\bibitem{Belitsky:2007cc}
  A.~V.~Belitsky,
  arXiv:0706.4121 [hep-th].

%
\end{thebibliography}
\end{document}